\documentclass{article}
\usepackage[table]{xcolor}
\usepackage{spconf,graphicx}
\usepackage[hyphens]{url}
\usepackage[]{hyperref}
\hypersetup{breaklinks=true}
\usepackage{spconf,amsmath,graphicx}
\usepackage{microtype}[activate]
\usepackage{cleveref}
\usepackage{balance}
\usepackage{multicol}
\usepackage{multirow}
\usepackage{booktabs}
\usepackage{xcolor}
\usepackage{soul}
\usepackage{graphicx}
\usepackage{subcaption}
\usepackage{comment}

\newcommand{\cf}{{cf.\,}}

\title{EIHW-MTG: Second DiCOVA Challenge System Report}
\name{Adria Mallol-Ragolta$^{1}$, Helena Cuesta$^{2}$, Emilia G\'omez$^{2,3}$, and Bj\"orn W.\ Schuller$^{1,4}$}
\address{\fontsize{11}{11}\selectfont
  $^{1}$ EIHW -- Chair of Embedded Intelligence for Health Care \& Wellbeing, University of Augsburg, Germany\\
  \fontsize{11}{11}\selectfont$^{2}$ MTG -- Music Technology Group, Universitat Pompeu Fabra, Spain\\
  \fontsize{11}{11}\selectfont$^{3}$ Joint Research Centre, European Commission, Spain\\
  \fontsize{11}{11}\selectfont$^{4}$ GLAM -- Group on Language, Audio \& Music, Imperial College London, UK}

\begin{document}
\ninept
\maketitle

\begin{abstract}
\noindent
This work presents an outer product-based approach to fuse the embedded representations generated from the spectrograms of cough, breath, and speech samples for the automatic detection of COVID-19. To extract deep learnt representations from the spectrograms, we compare the performance of a CNN trained from scratch and a ResNet18 architecture fine-tuned for the task at hand. Furthermore, we investigate whether the patients' sex and the use of contextual attention mechanisms is beneficial. Our experiments use the dataset released as part of the Second Diagnosing COVID-19 using Acoustics (DiCOVA) Challenge. The results suggest the suitability of fusing breath and speech information to detect COVID-19. An Area Under the Curve (AUC) of 84.06\,\% is obtained on the test partition when using a CNN trained from scratch with contextual attention mechanisms. When using the ResNet18 architecture for feature extraction, the baseline model scores the highest performance with an AUC of 84.26\,\%.
\end{abstract}
\section{Introduction}
\label{sec:introduction}

Digital health systems powered with \textit{Artificial Intelligence} (AI) have the potential to revolutionise the health care systems worldwide, %. These can provide highly scalable, cost-effective solutions to improve the early diagnosis of diseases, and the monitoring of the patients towards personalised treatment plans. 
improving the early diagnosis of diseases, and the monitoring of the patients towards personalised treatment plans. 
Previous works in the literature explored the use of AI-based techniques in a wide range of medical problems, including the detection of coughs or sneezes~\cite{Amiriparian17-CAD}, the analysis of breath signals~\cite{Schuller20-TI2}, or the recognition of mental illnesses, such as depression~\cite{Mallol19-AHA,Ringeval19-A2W} or Post-Traumatic Stress Disorder (PTSD)~\cite{Mallol18-AMA}. Such technologies do not aim at replacing medical diagnostic tools, rather providing highly scalable, cost-effective pre-screening solutions to optimise the medical resources. 

In the current pandemic context caused by the outbreak of the \textit{Coronavirus Disease 2019} (COVID-19), we envision the use of new technologies to help monitor the spread of this virus. As the COVID-19 symptomatology presents affections in the human respiratory system, it seems reasonable to argue about the potential of the respiratory-related sounds to contain salient information for the detection of this disease. Hence, there is an opportunity to develop new, digital solutions exploiting respiratory sounds to detect patients with COVID-19. 

% Brief literature review goes here

This work focuses on the automatic detection of patients with COVID-19 in the context of the Second \textit{Diagnosing COVID-19 using Acoustics} (DiCOVA) Challenge~\cite{Sharma20-CAD, Sharma22-TSD}. We use the spectrogram representation of cough, breath, and speech samples to train neural networks composed of two main blocks: the first block aims at extracting embedded representations from the spectrograms, the second block is responsible for the actual classification. The embedded representations from the different sound types are extracted with dedicated \textit{Convolutional Neural Networks} (CNNs). We explore the use of an outer product-based approach to fuse the extracted representations with the goal to enrich the information for the final classification. Additionally, we also aim to investigate whether using the patients' sex as \textit{a priori} information, and introducing contextual attention mechanisms to the network can be beneficial for the task at hand. 

%The rest of the paper is laid out as follows. \Cref{sec:dataset} describes the dataset analysed, while \Cref{sec:methodology} details the methodology followed. \Cref{sec:results} compiles and analyses the results obtained from the experiments performed, and \Cref{sec:conclusions} concludes the paper and suggests potential directions for future works. 

\section{System Description}

\subsection{Dataset}
\label{sec:system:dataset}
%The dataset used in this work was 
In this work, we use the dataset released as part of the Second DiCOVA Challenge~\cite{Sharma20-CAD, Sharma22-TSD}. This dataset contains acoustic samples of COVID-19 positive and negative (healthy) patients from three different sound types produced by the human respiratory system; specifically, from coughs, breaths, and speech. Although the sampling rate of the acoustic samples provided is 44.1\,kHz, an initial exploration of the dataset revealed the existence of samples without frequency content in the upper frequencies of the spectrogram. This observation suggests that some audio samples were originally recorded at a different, lower sampling rate, and upsampled before distributing the data. This is a plausible hypothesis given the nature of the dataset, which was recorded in-the-wild, via crowdsourcing, and using the patients' own devices. The available samples are distributed in two partitions, and the Challenge organisers require assessing the performance of the models on the training partition using a pre-defined 5-fold cross-validation approach. 

Each patient recorded a cough, a breathing, and a speech sample. The total duration of the dataset is 14\,h 45\,min 23\,sec (\cf \Cref{table:dataset_soundTypes}). The dataset contains information from a total of 1\,436 patients (\cf \Cref{table:dataset_participants}): 965 belonging to the training partition, and 471, to the test partition. The training data is imbalanced both in terms of sex (242 females and 723 males) and COVID-19 status (172 positives and 793 negatives). Similarly, the test data is also imbalanced in terms of sex (119 females and 352 males), whilst the COVID-19 status distribution is blind to the Challenge participants. 
\begin{table}[t]
%\caption{Summary of the data available in the Second DiCOVA Challenge dataset time-wise per sound type and data partition. The temporal information is provided in the format (HH):MM:SS.}
\caption{Data available in the Second DiCOVA Challenge dataset time-wise per sound type and data partition. The temporal information is provided in the format (HH):MM:SS.}
\label{table:dataset_soundTypes}
    \vspace{-1.5em}
    \begin{center}
    \resizebox{0.67\columnwidth}{!}{
        \begin{tabular}{lrrr}
            \toprule
             %& \multicolumn{1}{c}{\textbf{Validation}} & \multicolumn{1}{c}{\textbf{Test}} & \multicolumn{1}{c}{$\sum$}\\
             & \textbf{Validation} & 
             \textbf{Test} & 
             $\sum$\\
            \midrule
            Cough  &  1:41:01 &   37:58 &  2:18:59\\
            Breath &  4:37:37 & 2:07:46 &  6:45:23\\
            Speech &  3:56:22 & 1:44:39 &  5:41:01\\
            \midrule
            $\sum$ & 10:15:00 & 4:30:23 & 14:45:23\\
            \bottomrule
        \end{tabular}
    }
    %\vspace{-.5em}
    \end{center}
    \vspace{-1em}
\end{table}
\begin{comment}
\begin{table}[t]
%\caption{Statistics of the Second DiCOVA Challenge dataset in terms of the patients' sex and their COVID-19 status. The latter information on the test samples is blind to the Challenge participants. }
\caption{Statistics of the Second DiCOVA Challenge dataset in terms of the patients' sex and their COVID-19 status. The latter is blind to the Challenge participants on the test set. }
\label{table:dataset_participants}
\vspace{-1.5em}
\begin{center}
\resizebox{0.8\columnwidth}{!}{%
    \begin{tabular}{lrrrrr}
        \toprule
         & \multicolumn{3}{c}{\textbf{Validation}} & \multirow{2}{*}{\multicolumn{1}{c}{\vspace{-.5em}\textbf{Test}}} & \multirow{2}{*}{\multicolumn{1}{c}{\vspace{-.5em}$\sum$}} \\[2pt]
        \cline{2-4} 
        %\rule{0pt}{12pt} & \multicolumn{3}{c}{COVID-19 Status} & \multicolumn{1}{c}{\textbf{Test}} & \multicolumn{1}{c}{$\sum$} \\[2pt]
        %\cline{2-4} 
        \rule{0pt}{12pt} & Positive & Negative & $\sum$ & & \\[2pt]
        %\rule{0pt}{12pt} & Positive & Healthy & $\sum$ & & \\[2pt]
        \midrule
        Females   &  53 & 189 & 242 & 119 &    361\\
        Males     & 119 & 604 & 723 & 352 & 1\,075 \\
        \midrule
        $\sum$    & 172 & 793 & 965 & 471 & 1\,436 \\
      \bottomrule
    \end{tabular}
    }
\end{center}
\vspace{-2em}
\end{table}
\end{comment}
%
\begin{table}[t]
%\caption{Statistics of the Second DiCOVA Challenge dataset in terms of the patients' sex and their COVID-19 status. The latter information on the test samples is blind to the Challenge participants. }
\caption{Statistics of the Second DiCOVA Challenge dataset in terms of the patients' sex and their COVID-19 status. The latter is blind to the Challenge participants on the test set. }
\label{table:dataset_participants}
\vspace{-1.5em}
\begin{center}
\resizebox{0.8\columnwidth}{!}{%
    \begin{tabular}{lrrrrr}
        \toprule
         & \multicolumn{3}{c}{\textbf{Validation}} & \multirow{2}{*}{\textbf{Test}} & \multirow{2}{*}{$\sum$} \\[2pt]
        \cline{2-4} 
        %\rule{0pt}{12pt} & \multicolumn{3}{c}{COVID-19 Status} & \multicolumn{1}{c}{\textbf{Test}} & \multicolumn{1}{c}{$\sum$} \\[2pt]
        %\cline{2-4} 
        \rule{0pt}{12pt} & Positive & Negative & $\sum$ & & \\[2pt]
        %\rule{0pt}{12pt} & Positive & Healthy & $\sum$ & & \\[2pt]
        \midrule
        Females   &  53 & 189 & 242 & 119 &    361\\
        Males     & 119 & 604 & 723 & 352 & 1\,075 \\
        \midrule
        $\sum$    & 172 & 793 & 965 & 471 & 1\,436 \\
      \bottomrule
    \end{tabular}
    }
\end{center}
\vspace{-2em}
\end{table}
\subsection{Data Preparation}
\label{sec:system:dataPreparation}
The respiratory sounds are first downsampled to 16\,kHz to overcome the disparity between recording devices, avoiding our networks to perform the COVID-19 detection based on the presence or the absence of frequency content in the upper frequencies of the spectrogram (\cf \Cref{sec:system:dataset}). This work focuses on fusing the information embedded in the different sounds recorded by each patient. %These were recorded asynchronously and all have a different duration, which poses a challenge towards the fusion methods we aim to investigate. To overcome this issue, we compute the longest sound from each patient. We then use this information to repeat the shorter sounds, so that the cough, the breathing, and the speech samples from each patient have all the same duration. 
As each sound has a different duration, we compute the longest one from each patient and use this information to extend the shorter sounds via repetition, so all samples from each patient have the same duration. 
Next, we window each respiratory sound separately into frames of 5\,sec length with a 50\,\% overlap. We compute the magnitude of the \textit{Short-Time Fourier Transform} (STFT) of each individual frame using a window length of 4096 samples (256\,ms) and a hop size of 128 samples (8\,ms) to obtain its spectrogram representation. The spectrograms are generated using a logarithmic frequency scale, and the \textit{magma} colour map. Once normalised, each spectrogram is stored in disk as a colour image of $224 \times 224$ pixels. 

The generated spectrograms from each sound type are standardised before being fed into the models for training. The standardisation parameters ($\mu$ and $\sigma$) are computed from all the spectrograms corresponding to the current sound type that belong to the training partition. To downsize the effect of training the models with COVID-19 imbalanced data (\cf \Cref{table:dataset_participants}), we augment the generated spectrograms corresponding to the COVID-19 positive patients via replication to balance the training data. Despite considering other data augmentation strategies, such as filtering or additive noise, we decided not to alter the original samples in any way, as the relevant acoustic information for the task at hand is not clear yet. The replication approach may introduce redundancy in the training material; however, we believe it can still be useful in this case, as the number of positive and negative samples is significantly different. 
\subsection{Models Description}
\label{sec:system:modelsDescription}

This passage describes the network architectures implemented and investigated in this work.

\subsubsection{Baseline Models}
\label{sec:models_baseline}

The networks implemented are composed of two main blocks: the first block extracts deep learnt representations from the spectrograms of the cough, $\boldsymbol{f}_{C}$, breath, $\boldsymbol{f}_{B}$, and speech, $\boldsymbol{f}_{S}$, samples, while the second block performs the actual classification. For the feature extraction block, we compare two different architectures. The first architecture implements two convolutional layers with 16 and 4 filters, respectively, with a kernel size of $3 \times 3$ and a stride of 1. Following each convolutional layer, we use batch normalisation and the output is transformed using a \textit{Rectified Linear Unit} (ReLU) function. A 2-dimensional max pooling layer and a 2-dimensional adaptive average pooling layer are implemented at the end of the first and second convolutional block, respectively. This way, we force the output of the feature extraction block to produce 4 features per filter. The second architecture uses the ResNet18 architecture~\cite{He16-DRL} without the last layer. Specifically, we use the pre-trained weights to initialise the network and fine-tune them during training for the task at hand. An additional linear layer is included in this architecture to reduce the dimensionality of the features from 512 to 16. The learnt features from both architectures have the same dimensionality and are finally flattened into a 1-dimensional representation.

The deep learnt representations from each sound type are extracted using a dedicated feature extraction block. In this work, we investigate the inner fusion of these embedded representations using an outer product-based approach, which can be mathematically defined as:
\begin{equation}
    \boldsymbol{f}_{C \otimes B \otimes S}=\left[\begin{matrix} \boldsymbol{f}_{C} \\ 1
    \end{matrix}\right] \otimes \left[\begin{matrix}\boldsymbol{f}_{B} \\ 1\end{matrix}\right] \otimes \left[\begin{matrix}\boldsymbol{f}_{S}\\ 1\end{matrix}\right]. 
    \label{eq:outerProduct}
\end{equation}
When the three sound types are fused together, the outer product generates a cube with the following properties: i) the original representations are preserved in the edges of the cube, ii) each face of the cube contains information from the fusion of 2 sound types, and iii) the inner part of the cube fuses information from the three sound types all together. The fused representation is flattened before being fed into the final, classification block of the network. This fusion layer is implemented when training multi-type models, which combine at least two sound types, and omitted when training mono-type models, which consider a single sound type to infer the COVID-19 status.

The classification block of the network contains two fully connected layers, preceded by a dropout layer with probability 0.3. The number of input neurons in this block depends on the number of sound types selected for training. Nevertheless, the number of output neurons is fixed to 8. The output of this first layer is transformed using a ReLU activation function. The transformed representation is finally fed into the second layer of this block, which contains two output neurons with a Softmax activation function. This way the outputs of the network can be interpreted as probability scores. 

\subsubsection{Sex-based Models}
\label{sec:models_sex}

This model expands the baseline model described in \Cref{sec:models_baseline} to consider the sex of the patients when inferring their COVID-19 status. Specifically, a binary encoded representation of the patient's sex is fed into the second layer of the classification block of the network. The number of input features to the classification block depends on the number of sound types to be fused. Introducing the sex information in the first layer of this block would difficult understanding if the performance of the network is conditioned by the patient's sex or by the number of input features. Thus, we opted for feeding this information into the second layer of the classification block, where the number of neurons corresponding to the sound representations is fixed. 

\begin{table}[t]
%\caption{Summary of the Area Under the Curve (AUC) measurements, in percentage, obtained from the mono-type and multi-type models trained using a dedicated CNN-based network (Baseline). These models are studied in three different scenarios: considering the patient's sex to perform the analysis (Sex), using contextual attention mechanisms (C. Att.), and combining patients' sex and contextual attention mechanisms (Sex \& C. Att.).}
\caption{AUC measurements (\%) obtained from the mono- and multi-type models trained using a dedicated CNN-based network (Baseline). These models consider the patient's sex for the analysis (Sex), use contextual attention mechanisms (C. Att.), and their combination (Sex \& C. Att.).}
\label{table:scratch_results}
\vspace{-1.5em}
\begin{center}
\resizebox{\columnwidth}{!}{
    \begin{tabular}{lrrrrr}
    \toprule
    % header
    %\multirow{2}{*}{\textbf{Sound types}} & 
    %\multirow{2}{*}{\multicolumn{1}{c}{\textbf{Sound types}}} & 
    \multirow{2}{*}{\textbf{Sound types}} & 
    %\multirow{2}{*}{\multicolumn{1}{c}{\textbf{Set}}} &
    \multirow{2}{*}{\textbf{Set}} &
    %\multirow{2}{*}{\multicolumn{1}{c}{\textbf{Baseline}}} & 
    \multirow{2}{*}{\textbf{Baseline}} & 
    %\multirow{2}{*}{\multicolumn{1}{c}{\textbf{Sex}}} & 
    \multirow{2}{*}{\textbf{Sex}} & 
    %\multirow{2}{*}{\multicolumn{1}{c}{\textbf{C. Att.}}} & 
    \multirow{2}{*}{\textbf{C. Att.}} & 
    \multicolumn{1}{c}{\textbf{Sex \&}} \\
    & & & & & \multicolumn{1}{c}{\textbf{C. Att.}} \\ \midrule
    % end header
    \multirow{2}{*}{$C$} & Val. & 63.56 & 65.16 & 63.86 & 67.62 \\
    & Test & 61.56 & 65.16 & 64.01 & 67.71\\ \cmidrule{1-6}
    \multirow{2}{*}{$B$} & Val. & 72.83 & 73.83 & 72.73 & 71.96 \\
    & Test & 79.38 & 79.85 & 76.51 & 76.79\\ \cmidrule{1-6}
    \multirow{2}{*}{$S$} & Val. & 71.90 & 72.92 & 72.49 & 73.52\\
    & Test & 80.04 & 75.53 & 78.35 & 78.32\\ \cmidrule{1-6}
    \multirow{2}{*}{$C \otimes B$} & Val. & 74.68 & 74.94 & 74.14 & 75.41\\
    & Test & 80.02 & 80.37 & -- & -- \\ \cmidrule{1-6}
    \multirow{2}{*}{$C \otimes S$} & Val. & 72.45 & 71.59 & 74.58 & 74.40\\
    & Test & -- & -- & -- & --\\ \cmidrule{1-6}
    \multirow{2}{*}{$B \otimes S$} & Val. & 76.74 & 77.98 & 76.04 & 77.92\\
    & Test & \textbf{81.95} & \textbf{83.89} & \textbf{84.06} & 82.35\\ \cmidrule{1-6}
    \multirow{2}{*}{$C \otimes B \otimes S$} & Val. & 72.91 & 73.71 & 78.22 & 76.77 \\
    & Test & -- & -- & 81.98 &  \textbf{83.25}\\
    \bottomrule
    \end{tabular}
}
\end{center}
\vspace{-2em}
\end{table}

\subsubsection{Contextual Attention-based Models}
\label{sec:models_attention}

This model also expands the baseline model described in \Cref{sec:models_baseline}, but, in this case, using a dedicated contextual attention mechanism at the output of each feature extraction block. The aim of this mechanism is to help highlight the salient information from the embedded representations learnt. Representing the embedded representations learnt as $\boldsymbol{f}_{N}$, where $N = {C, B, S}$ depending on the input sound type, the contextual attention mechanism is mathematically defined as:
\begin{equation}
    \boldsymbol{u} = \tanh(\mathbf{W} \boldsymbol{f}_N + \mathbf{b}),
\end{equation}
\begin{equation}
    \boldsymbol{\alpha} = \frac{\exp\left(\boldsymbol{u}^T \mathbf{u_c}\right)}{\sum \exp\left(\boldsymbol{u}^T \mathbf{u_c}\right)},
\end{equation}
\begin{equation}
    \boldsymbol{\tilde{f}}_N = \boldsymbol{\alpha} \boldsymbol{f}_N, 
\end{equation}
where $\mathbf{W}$, $\mathbf{b}$, and $\mathbf{u_c}$ are parameters to be learnt by the network. The parameter $\mathbf{u_c}$ can be interpreted as the context vector. The attention-based representation obtained, $\boldsymbol{\tilde{f}}_N$, is then fed into the classification block of the network when training mono-type models, or fused when training multi-type models. 

\subsection{Networks Training}
\label{sec:system:netTraining}

For a fair comparison among the models, these are all trained under the exact same conditions. We use the Categorical Cross-Entropy as the loss to minimise, using Adam as the optimiser with a fixed learning rate of $10^{-3}$. As model performances are assessed in terms of the \textit{Area Under the Curve} (AUC), we define $\mathcal{L}_{AUC} = 1 - AUC$ as the validation loss to monitor during the training process. Network parameters are updated in batches of 64 samples, and trained during a maximum of 100 epochs. We implement an early-stopping mechanism to stop training when the validation loss does not improve for 15 consecutive epochs. We follow a 5-fold cross-validation approach to evaluate the models, as defined by the Challenge organisers. Each fold is trained during a specific number of epochs. Hence, when modelling all training material and to prevent overfitting, the training epochs are determined by computing the mean of the training epochs processed in each fold, rounded up to the next integer. 

\section{Experimental Results}
\label{sec:results}

\begin{table}[t]
%\caption{Summary of the Area Under the Curve (AUC) measurements, in percentage, obtained from the mono-type and multi-type models trained using a ResNet18 Architecture-based network (Baseline). These models are studied in three different scenarios: considering the patient's sex to perform the analysis (Sex), using contextual attention mechanisms (C. Att.), and combining patients' sex and contextual attention mechanisms (Sex \& C. Att.).}
\caption{AUC measurements (\%) obtained from the mono- and multi-type models trained using a ResNet18-based network (Baseline). These models consider the patient's sex for the analysis (Sex), use contextual attention mechanisms (C. Att.), and their combination (Sex \& C. Att.).}
\label{table:resnet_results}
\vspace{-1.5em}
\begin{center}
\resizebox{\columnwidth}{!}{
    \begin{tabular}{lrrrrr}
    \toprule
    % header
    %\multirow{2}{*}{\textbf{Sound types}} & 
    %\multirow{2}{*}{\multicolumn{1}{c}{\textbf{Sound types}}} & 
    \multirow{2}{*}{\textbf{Sound types}} & 
    %\multirow{2}{*}{\multicolumn{1}{c}{\textbf{Set}}} &
    \multirow{2}{*}{\textbf{Set}} &
    %\multirow{2}{*}{\multicolumn{1}{c}{\textbf{Baseline}}} & 
    \multirow{2}{*}{\textbf{Baseline}} & 
    %\multirow{2}{*}{\multicolumn{1}{c}{\textbf{Sex}}} & 
    \multirow{2}{*}{\textbf{Sex}} & 
    %\multirow{2}{*}{\multicolumn{1}{c}{\textbf{C. Att.}}} & 
    \multirow{2}{*}{\textbf{C. Att.}} & 
    \multicolumn{1}{c}{\textbf{Sex \&}} \\
    & & & & & \multicolumn{1}{c}{\textbf{C. Att.}} \\ \midrule
    % end header
    \multirow{2}{*}{$C$} & Val. & 76.42 & 74.48  & 73.12 & 73.39  \\
    & Test & 64.69 & 68.76 & 66.60 & 68.15\\ \cmidrule{1-6}
    \multirow{2}{*}{$B$} & Val. & 78.78  & 79.16 & 78.62 &  80.78 \\
    & Test & 80.35 & \textbf{79.91} & 77.77 & 80.21\\ \cmidrule{1-6}
    \multirow{2}{*}{$S$} & Val. & 79.02 & 79.25 & 78.19 & 79.10 \\
    & Test & 81.86 & 75.21 & 78.89 & \textbf{81.66}\\ \cmidrule{1-6}
    \multirow{2}{*}{$C \otimes B$} & Val. & 76.35 & 76.79 & 74.46 & 72.48\\
    & Test & -- & 75.03 & -- & --\\ \cmidrule{1-6}
    \multirow{2}{*}{$C \otimes S$} & Val. & 75.56 & 76.19 & 77.69 & 76.64\\
    & Test & -- & -- & 77.07 & --\\ \cmidrule{1-6}
    \multirow{2}{*}{$B \otimes S$} & Val. & 78.06 & 79.87 & 80.12 & 77.65 \\
    & Test & \textbf{84.26} & 73.48 & \textbf{83.63} & 81.48\\ \cmidrule{1-6}
    \multirow{2}{*}{$C \otimes B \otimes S$} & Val. & 76.90 & 75.84 & 76.79 & 76.07 \\
    & Test & 76.78 & -- & -- & --\\
    \bottomrule
    \end{tabular}
}
\end{center}
\vspace{-2em}
\end{table}

The results obtained %assessing the models trained 
using specific CNNs and using %pre-trained ResNet18 to initialise the network weights 
ResNet18-based CNNs are summarised in Tables~\ref{table:scratch_results}~and~\ref{table:resnet_results}, respectively.
%One of the main insights from our experiments is that the fusion of breath and speech samples outperforms all other sound type combinations in all but one case for the CNN-based network, and two out of four cases for the ResNet18-based network.
One of the main insights from our experiments is that the fusion of breath and speech samples outperforms the multi-type models resulting from the combination of all other sound types and the mono-type models %in 3 out of the 4 scenarios investigated with the dedicated CNN, and in 2 out of the 4 scenarios compared with the ResNet18-based CNN. 
in 3 out of the 4, and in 2 out of the 4 scenarios investigated with the specific CNNs, and the ResNet18-based CNNs, respectively.
%Likewise, when we look at the mono-type models ($C$, $B$, $S$), we observe that the breath and speech modalities score significantly higher results than the cough in both network architectures.
Likewise, when we look at the mono-type models ($C$, $B$, $S$), we observe that the models using the breath and the speech samples score %significantly
higher results in comparison to the models using coughs only.%only the cough information. % in both network architectures.
%
%
%\hc{Interestingly, in the CNN-based scenario, considering the sex of the patient has a negative impact on the speech type, while it seems to be beneficial in the case of coughs, and to barely have any effect in the case of breaths.
%If we compare the results between the CNN- and ResNet18-based networks, we can conclude that using transfer learning is beneficial for this task. Looking at the fusion of breaths and speech, which obtains better results overall, we observe a slightly superior performance by some of the ResNet18-based networks, \ie, the ResNet18-based baseline scores 2.3\,\% more than its CNN-based counterpart.
%In the multi-type models, we see that differences between the baseline models and their three variants are rather small. Hence, more experiments looking into the outputs of the models would help in the assessment of the effect of adding the patient's sex and using contextual attention.
%Finally, comparing Tables~\ref{table:scratch_results}~and~\ref{table:resnet_results} we also observe that the fusion of two or more sound types is more effective with the CNN-based networks, and we find the fusion of the three sound types to obtain the best performance in only one of the eight studied scenarios. }

We observe that the mono-type models considering the patients' sex only improve the performance of the cough-based models, while they barely have an effect on the breath-based models. Patients' sex negatively impacts the performance of the speech-based models.
Although there is no clear pattern to determine the suitability of considering patients' sex and/or using contextual attention, we note that the models surpassing the baseline with the specific CNNs use one of the three variants in most of the cases. The contextual attention-based model fusing breath and speech samples obtains the best performance with an AUC of 84.06\,\%. With the ResNet18-based CNNs, the baseline models obtain the best AUC scores in most of the cases. The baseline model fusing breath and speech samples scores the best AUC of 84.26\,\%.

Although the transfer learning approach obtains the best performance, the specific CNNs obtain similar results with a simpler structure. Further experiments are needed to better understand the impact of patients' sex in the fused scenarios, as we hypothesise it is downsized as a result of a magnitude difference between the sex representation and the deep learnt features at the intermediate layer of the classification block. 

%\section{Conclusions}
%\label{sec:conclusions}

\section{Acknowledgements}
This project has received funding from the European Union's Horizon 2020 research and innovation programme under grant agreement No.\,826506 (sustAGE), and from the Spanish Ministry of Science and Innovation under the Musical AI project (PID2019-111403GB-I00).
\bibliographystyle{IEEEtran}

\bibliography{mybib}

\end{document}